\documentstyle[12pt]{article}

\def\Tr{\mathop{\rm Tr}}

\begin{document}

\begin{titlepage}

\def\thefootnote{\ast}

\begin{flushright}
ISC--94--1 \\[2pt]
April 1994
\end{flushright}
\vspace{40pt}

\begin{center}
{\Large\bf
Quantum reparametrizations \\
in the two-dimensional gravity: \\[10pt]
a look from 2+$\epsilon$ dimensions } \\[40pt]

S.~N{\sc aftulin}
\footnote{ Electronic address: \ naftulin@isc.kharkov.ua}
\\[7pt]

{\it  Institute for Single Crystals,  \\[-2pt]
60 Lenin Ave., 310141 Kharkov, Ukraine}  \\
\end{center}
\bigskip

\begin{abstract}
We discuss the structure of one-loop counterterms for the two-dimensional
theory of gravitation in the covariant scheme and study the effect of
quantum reparametrizations. Some of them are shown to be equivalent
to the introduction of $2+\epsilon$-dimensional terms into the initially
2-dimensional theory. We also argue that the $\beta$-function for the
Einstein constant has a non-trivial ultraviolet stable point beyond two
dimensions.
\end{abstract}

\end{titlepage}

\def\thefootnote{\arabic{footnote}}
\setcounter{footnote}{0}

\section{Introduction}

The celebrated $\epsilon$-expansion devised primarily for the needs of
critical phenomena (see \cite{Kogut_Wilson}
and references therein) has paved its way into the quantum theory of
gravitation \cite{E-expansion_in_gravity,Einstein_gravity_in_d=2+E}.
The approach received a new impact when it was realized that it pays to start
from $d=2+\epsilon$ and then analytically continue $\epsilon\to2\,$.
(The odd-dimensional theories are generally excluded from such an analytic
continuation.)

Basically, some of geometrical terms may drop out as the space-time dimension
decreases by an integer. However, at non-integer values of $d$ there is no
independent Lagrangian construction, and the theory is defined by an analytic
continuation in the parameter space. Thus one can expect that a smooth
reduction of $d$ (in the spirit of the renormalization group) leaves
its ``fingerprints'', in the reparametrization structure of the resulting
theory: the dropped structures degenerate and mix with those left. An obvious
example is two actions quadratic in curvature,
$S_1=\smallint\!d^2x\sqrt{g}R^{\mu\nu}R_{\mu\nu}$ and
$S_2={1\over2}\smallint\!d^2x\sqrt{g}R^2\,$, which become indistinguishable
in two dimensions, by the Bianchi identities:
\begin{equation}
R_{\mu \nu }={1\over2}Rg_{\mu\nu} \ .
\label{Bianchi_identities}
\end{equation}
The first impulse is to discard $S_1$ as we descend to a $d=2$ theory, but
this would simply entail the impossibility of returning to higher dimensions.
A more sophisticated option is to allow for an operator mixing; the structure
degenerates as we approach  $d=2$ and the degeneracy should be reflected in
the invariance of the resulting theory with respect to possible
parametrizations. Conversely, as we perform an analytic continuation beyond
$d=2$ the degeneracy, (\ref{Bianchi_identities}), is lifted and additional
terms come into effect. Hence the study of different quantum parametrizations
is not completely meaningless,%
\footnote{We do not touch upon reparametrization anomalies here.}
although the on-shell effective actions are parametrization independent
up to topologically trivial surface terms
\cite{equivalence_theorem,background_field_method}.

Another relevant problem is the presence of ultraviolet divergences in any
integer space-time dimension. There are several covariant regularizations, of
which the most suitable for our purposes is the dimensional reduction:
$d=2\to2(1-\epsilon)\,$, $\,\epsilon\ge0\,$. Although there are no
singularities for intermediate values of $\epsilon\,$, the limit
$\epsilon\to0$ is special at each order of the perturbation theory in the
Einstein constant $\kappa\,$. For small but finite values of $\,\epsilon\,$
a vast number of terms arises in the effective action; however, the would-be
convergent terms contribute an extra power of $\epsilon\,$, which is
negligibly small within the loop expansion. We will restrict our attention
to the one-loop approximation and small $\epsilon\,$. Consequently the
quantum corrections which are finite in the limit $\epsilon\to0\,$, will be
of no interest to us.

It has been known for some time that the treatment of the $d=2$ Einstein
theory based upon the dimensional regularization is afflicted by
oversubtractions, \cite{Einstein_gravity_in_d=2+E},
since the action $\smallint\!d^2x\sqrt{g}R$ is a topological invariant. Due
to peculiar features of the conformal mode at $\epsilon\to0$ there seems to
be a better candidate for a $2+\epsilon$-dimensional theory of gravitation,
viz., the Jackiw-Teitelboim model \cite{JT-model}.
It is very likely that this model does not admit the problem of
oversubtractions at $\epsilon\to0$ because the dilaton, $\Phi\,$, field may
be viewed as arising from the integration over the $d=2$ conformal anomaly,
\cite{dilaton_gravity_from_anomaly}.
The interplay between the dilaton and the conformal mode should result in a
cancellation of the kinematical ${\epsilon}^{-1}$-poles in the propagators.

The dimensional regularization scheme has another important advantage:
it automatically eliminates the contributions like $\delta(0)$ to the
effective action. Due to this property the dimensional regularization
ensures consistency of the naive determinant calculations, which would
otherwise require a modification when applied to gravity
\cite{Barvinsky-Vilkovisky:review}.

At the first sight, the conformal properties of two-dimensional models of
gravity suggest the natural, conformal, parametrization of the quantum metric
fluctuations $g_{\mu\nu}\to\widetilde{g}_{\mu\nu}=(\exp\sigma)g_{\mu\nu}\,$
(we use the language of the background field method
\cite{background_field_method}
throughout the paper, also see a book \cite{Buchbinder_etal:book}
for a comprehensive review). But on the other hand, the conformal
representation does not seem very suitable in view of a possible extension to
the
higher-dimensional world ($\epsilon\to2$). For example, the conformal
approach to the $d=4$ gravity
\cite{conformal_factor_dynamics_at_d=4}
has attained a limited success as the nice properties of exact solvability
are destroyed. Thus it may be crucial to see what the picture looks like
\cite{Odintsov_Shapiro}
in the conventional, linear, representation
$g_{\mu\nu}\to\widetilde{g}_{\mu\nu}=g_{\mu\nu}+h_{\mu\nu}\,$.

In this paper, we re-examine the structure of one-loop counterterms in
two-dimensional gravities and make a brief glance at how the
reparametrization effects modify the considerations. Although the main
logical emphasis might be on the conventional gravity, we find it more
convenient to start (Section~2) with its higher-order counterpart. Section~3
contains a paralleling treatment of the dilaton gravity, and a short
Section~4 is devoted to conclusions.

\section{Resurrecting Gravitons in the $R^2$-gravity}

The two-dimensional $R^2$ quantum gravity does not apparently have any
independent physical significance and mainly serves as a good playground.
It was extensively studied within the Arnovitt-Deser-Misner formalism and was
found to have zero propagating degrees of freedom \cite{Yoneya}
and hence a trivial $S$-matrix.%
\footnote{The same feature is shared by the Jackiw-Teitelboim model and its
generalizations.}
Of course, the latter does not imply that the model describes trivial
space-time manifolds, nor that its ultraviolet divergences are absent.

Consider the linear background versus quantum metric splitting
$g_{\mu\nu}\to\widetilde{g}_{\mu\nu}=g_{\mu\nu}+h_{\mu\nu}\,$ and further
decompose the quantum fluctuation into its trace $h=g^{\mu\nu}h_{\mu\nu}$
and the traceless ``transverse graviton''
$\bar{h}_{\mu\nu}=h_{\mu\nu}-(1/2)hg_{\mu\nu}\,$. Since there are no
transverse directions in the $(1+1)$-dimensional space-time, it is natural to
assume that $\bar{h}_{\mu\nu}$ is just a gauge artifact; and indeed, one can
adopt the gauge $\bar{h}_{\mu\nu}=0\,$, which is essentially equivalent to
the background conformal gauge
$h_{\mu\nu}=\left(e^\sigma-1\right)g_{\mu\nu}\,$.

In the fourth-order theory, the number of degrees of freedom is effectively
doubled \cite{Stelle}
as compared to the conventional (second-order) one, so we can expect that
beside $h$ another dynamical scalar exists. This is at the heart of the
approach suggested by Yoneya \cite{Yoneya}:
using an auxiliary scalar field $\Phi$ the fourth order Lagrangian
\begin{equation}
S=-\int\!d^2x\,\sqrt g\,\left(\omega R^2+\Lambda\right)
\label{action_for_R2-gravity}
\end{equation}
may be represented as
\begin{equation}
S=-\int\!d^2x\,\sqrt g\,\left(R\Phi-{1\over4\omega}\Phi^2+\Lambda \right) \ .
\label{R2_action_with_aux_field}
\end{equation}
In the above definition, (\ref{action_for_R2-gravity}), we have neglected a
topological term
\begin{equation}
-{1\over2\kappa^2}\int\!d^2x\,\sqrt g\,R \ ,
\label{topological_term}
\end{equation}
because its appearance does not affect the divergent structure and only leads
to a constant shift of the ``dilaton'' field: $\,\Phi\to\Phi+(1/2\kappa^2)\,$.

Thus the problem of finding the one-loop structure of counterterms reduces to
that for dilaton gravities. The latter can be solved by a variety of methods,
either conformal or covariant. In the background field formulation, the
divergent contribution to the one-loop effective action, $\Gamma$, is,
\cite{our_appendix_on_R2_gravity}:
\begin{equation}
\Gamma_{div}=-{1\over4\pi\epsilon}\int\!d^2x\,\sqrt{g}\,R
\label{Gamma_div_for_R2-gravity}
\end{equation}
plus curvature terms at the one-dimensional boundary; here $\epsilon=(2-d)/2$
is the dimensional regulator.

The above result is not totally unexpected:
Eq.(\ref{Gamma_div_for_R2-gravity}) is the only metric-covariant local
expression with the appropriate background dimensionality. Since
$\smallint\!d^2x\sqrt{g}R$ is a topological invariant
Eq.(\ref{Gamma_div_for_R2-gravity}) is consistent with the anticipation that
the $S$-matrix is finite. To summarize, the model
(\ref{action_for_R2-gravity}) is renormalizable in the generalized sense
(i.e., after inclusion of the topological term (\ref{topological_term})
into the bare action), the cosmological constant, $\Lambda\,$, remains
finite.

Thus far, the treatment has been straightforward and it is not at all
evident why bother studying the same model in the covariant scheme.
(One such motivation might be to include the term
$\smallint\!d^2x\sqrt{g}R^{\mu\nu}R_{\mu\nu}$ into the bare action
(\ref{action_for_R2-gravity}) in the vicinity of $d=2\,$, see below). Even
the first step of the evaluation brings about surprises, in the guise of
technical obstacles. Thus it is worth while to say a few words of the
Schwinger-DeWitt technique (see, e.g., \cite{Barvinsky-Vilkovisky:review}
for a general review, and the second paper in \cite{Odintsov_Shapiro}
for applications to the two-dimensional gravity.)

The whole procedure reduces to the calculation of the determinant of the
fourth-order differential operator $\widehat{H}_{ij}\,$, which is essentially
the second functional derivative of the action (\ref{action_for_R2-gravity}).
This operator contains $2\times2$-matrices acting in the space of the
quantum fields $\{h;\,\bar{h}_{\mu\nu}\}\,$. If minimal gauge conditions are
used, it takes the form
\begin{equation}
\widehat{H}_{ij}=-\widehat{K}_{ij}\Delta^2+\mbox{lower-order derivatives }.
\label{min_operator}
\end{equation}
The divergences are defined by the coincidence limits in the heat kernel
expansion, or equivalently, through the ``universal functional traces''
\cite{Barvinsky-Vilkovisky:review}
obtained by iterating Eq.(\ref{min_operator}) with respect to its
highest-order term $\widehat{K}_{ij}\Delta^2\,$. Congenially, the symmetric
matrix $\widehat{K}_{ij}$ is usually taken to be a metric in the
configuration space of the quantum fields and hence defines the quantum
measure in the path integral, \cite{def_unique_eff_action}.

Quite curiously, expanding the basic action (\ref{action_for_R2-gravity})
in powers of the quantum fields one verifies that there appears no term like
$\,\bar{h}^{\mu\nu}\Delta^2\bar{h}_{\mu\nu}\,$ so that the matrix
$\widehat{K}_{ij}$ in Eq.(\ref{min_operator}) is degenerate, thus reminding
that the space of states is full of gauge phantoms---and that the field
$\bar{h}_{\mu\nu}$ is just one of them. Indeed, in higher dimensions the
corresponding term comes from the Weyl tensor squared (see, e.g.,
\cite{Stelle}),
which identically vanishes for $d=2\,$. The subsequent analysis may be
twofold: one could either gauge the field $\bar{h}_{\mu\nu}$ away adopting
the conformal parametrization, or rather invent some way to change the
relative weight of the conformal mode in the $h_{\mu\nu}$-loop by hand. It
can be argued \cite{our_R2_dilaton_gravity}
that the both procedures may be fitted so as to give equivalent off-shell
expressions for the one-loop divergences (up to surface terms). Here we only
take the second option.

In order to modify the conformal mode at the quantum level without breaking
the general co-ordinate covariance, let us consider the following term:%
\footnote{To our knowledge, a similar term was firstly introduced in
Ref.\cite{Kantowski_Marzban}
in a somewhat different setting.}
\begin{equation}
\delta S=-\xi\omega\int\!d^2x\,\sqrt{g}\bar{h}^{\mu\nu}\Delta\left[
          \widetilde{R}_{\mu\nu}-{1\over2}\widetilde{R}\widetilde{g}_{\mu\nu}
          \right] \ .
\label{delta_S}
\end{equation}
Here the quantities with tildes contain both background and quantum
components and hence must be Taylor expanded (to the first order in
fluctuations). The weight factor $\xi$ is arbitrary: as the expression in the
square brackets on the right-hand side of Eq.(\ref{delta_S}) is zero at
exactly $d=2$ by the virtue of the Bianchi identities
(\ref{Bianchi_identities}),
the divergent part of the effective action obviously does not depend on
$\xi\,$.%
\footnote{Rigorously, renormalization of the background fields might be
required to eliminate the reference to $\xi$ in the effective action.}

Adding Eq.(\ref{delta_S}) to the initial action (\ref{action_for_R2-gravity})
is equivalent to a reparametrization of the quantum fields so that the
divergences are not affected. However, one cannot send $\xi$ to zero until
the evaluation is complete because the matrix $\widehat{K}^{-1}$ gets
diverged. Further, the total contribution to the path integral measure,
$(i/2)\log\det\left(-\widehat{K}_{ij}\right)\,$, is proportional to
$\delta(0)\times\log\xi\,$, that is why the use of the dimensional
regularization is preferable: then $\delta(0)$ is regulated to zero. The
intermediate expressions acquiring a pole at $\xi=-1$, we confine ourselves
to the domain $\xi>0\,$.

Unfortunately, it is not evident either if the ghost operator becomes both
minimal and non-degenerate for $\xi\ne0\,$: this pivotal point must be
checked explicitly. As typical in the higher-order gauge theories, the gauge
fixing action,
\begin{equation}
S_{g.f.}=-\int\!d^2x\,\sqrt{g}\,\chi^\mu\widehat{C}_{\mu\nu}\chi^\nu \ ,
\label{gauge_fixing}
\end{equation}
contains the operator-valued $\widehat{C}_{\mu\nu}$ so that the third ghost
should be accounted for. A relatively simple choice of gauge is:
\begin{equation}
\chi^\mu=-\nabla_\nu\bar{h}^{\mu\nu}+{1\over2(1+\xi)}\nabla^\mu h \ ,
\end{equation}
\begin{equation}
\widehat{C}_{\mu\nu}=\xi g_{\mu\nu}\Delta+\nabla_\mu\nabla_\nu-\xi R_{\mu\nu}
                                                                       \ .
\label{third_ghost_operator}
\end{equation}
The total contribution to the one-loop effective action is given by the
standard expression
\begin{equation}
\Gamma={i\over2}\Tr\log\widehat{H}-i\Tr\log\widehat{\cal M}+{i\over2}\Tr\log
       \widehat{C} \ ,
\label{structure_of_Gamma_div}
\end{equation}
where the first term is determined by the quadratic expansion of
Eqs.(\ref{action_for_R2-gravity}), (\ref{delta_S}):
\begin{equation}
\widehat{H}_{ij}=\left(S^{(2)}+\delta S^{(2)}+S_{g.f.} \right)_{ij} \quad ,
\end{equation}
the second term is the Faddeev-Popov ghost operator:
\begin{eqnarray}
\widehat{\cal M}_{\mu\nu}&\equiv&\widehat{C}_{\mu\lambda}{\delta\chi^\lambda
                                  \over\delta\omega^\nu}      \nonumber\\
                        &=&\xi\Delta^2g_{\mu\nu}+R\nabla_\mu\nabla_\nu+
\mbox{lower-order derivatives },
\label{FP-ghost_operator}
\end{eqnarray}
and the third one is due to Eq.(\ref{third_ghost_operator}). Infinitesimal
parameters of the gauge transformations in Eq.(\ref{FP-ghost_operator}) are
denoted by $\,\omega^{\nu}$.

The last term in Eq.(\ref{structure_of_Gamma_div}) coincides (up to the
contribution to the functional measure) with the one-loop QED determinant
in the background Lorentz gauge with the parameter $\lambda=-1/(1+\xi)\,$.
In two space-time dimensions the divergent part of the QED is a
$\lambda$-independent \cite{Barvinsky-Vilkovisky:review}
surface term, though it is well-defined only for $\lambda>-1\,$, i.e., for
$\xi>0\,$.

With the account of the gauge fixing term (\ref{gauge_fixing}), the
$\xi$-dependence penetrates into the denominators of the configuration space
matrices in $\widehat{H}_{ij}\,$. The Vilkovisky-DeWitt metric with the
``resurrected'' graviton becomes
\begin{equation}
\widehat{K}={\omega\over2}\pmatrix{\xi/(1+\xi) & 0 \cr 0 &
             -\xi\left(g^{\mu\alpha}g^{\nu\beta}+g^{\mu\beta}g^{\nu\alpha}
             -g^{\mu\nu}g^{\alpha\beta}\right) \cr} \ .
\label{VD-metric}
\end{equation}
Note that the $hh$-propagator also gets modified, viz., $1\to\xi/(1+\xi)\,$
in its numerator. Now there is a simple way to see how the
$\,\bar{h}_{\mu\nu}\bar{h}_{\alpha\beta}$-sector decouples: set
$\xi\to\infty\,$ for instance.

In Ref.\cite{our_R2_dilaton_gravity}
we have demonstrated that all the $\xi$-dependent terms cancel exactly from
the final expression for $\Gamma_{div}\,$. This is a pleasant surprise since
in a general case renormalization of the metric $g_{\mu\nu}\,$ is needed to
eliminate $\xi$ from the effective action. We conclude that the divergences
calculated in this way coincide with the conformal gauge treatment (up to
curvature terms at the one-dimensional boundary).

The principal question here is: What is the nature of $\delta S\,$? In
Ref.\cite{Kantowski_Marzban} it was noticed that the $\xi$-dependence
represents parametrization ambiguities of the effective action (and hence
must vanish on shell), although their specific realization was not clarified.
The origin of this quantum reparametrization has nothing to do with gauge
fixing because $\xi$ had entered the action before a particular gauge was
imposed. The problem is not likely to be resolved in the formalism of the
Vilkovisky-DeWitt unique effective action \cite{def_unique_eff_action}:
its ``uniqueness'' does not rule out the possibility of having results that
depend on the choice of the configuration-space metric, and the preferred
metric (\ref{VD-metric}) contains $\xi$ explicitly.

Another related issue is that $\delta S$ is identically zero at $d=2$ since
the Bianchi identities hold for arbitrarily large metric disturbances (i.e.,
$\widetilde{R}_{\mu\nu}\equiv(1/2)\widetilde{R}\widetilde{g}_{\mu\nu}\,$ in
our notations); then, how can one obtain non-trivial contributions just by
expanding (\ref{Bianchi_identities})? The question may be re-formulated as
follows: What action corresponds to $\delta S$ at the tree level? There seems
to be no appropriate action functional at $d=2\,$.

In view of the degeneracy alluded to above, the natural place to try is
$2+\epsilon$ dimensions rather than exactly two: there must appear new
contributions like $R_{\mu\nu}R^{\mu\nu}$ or
$R_{\mu\nu\alpha\beta}R^{\mu\nu\alpha\beta}$. (We simply ignore the latter
aiming at the continuation to $d=4\,$.) To keep up with the dimensions an
arbitrary unit of mass $\mu\,$ should be introduced. Consider the following
term
\begin{equation}
{\cal A}=2\xi{\mu}^{2\epsilon}\int\!d^dx\,\sqrt g\,\omega\left(R_{\mu\nu}
                                        R^{\mu\nu}-{1\over2}R^2\right) \ .
\label{2+E_correction_to_R2}
\end{equation}
If we expand (\ref{2+E_correction_to_R2}) in powers of the quantum fields
$\{h;\,\bar{h}_{\mu\nu}\}$ and {\it after this\/} take the limit
$\epsilon\to0\,$, then $\delta S^{(2)}$ is immediately reproduced. (It is
important to realize that taking the limit $\epsilon\to0$ does not commute
with splitting the metric into its background and quantum parts.) There is
no extra contribution to the first order in fluctuations (i.e.,
$\,{\cal A}^{(1)}=0\,$) and the equations of motion remain unmodified. In
other words, one studies a theory in which there are $2+\epsilon$-dimensional
quantum fluctuations around the purely two-dimensional background. Let us
emphasize that the quantum reparametrization is equivalent to adding the term
$\smallint\!d^2x\sqrt{g}R_{\mu\nu}R^{\mu\nu}\,$ into the initial action
(\ref{action_for_R2-gravity}).

The limit $\xi\to\infty$ regains the corresponding one-loop determinant in
the conformal gauge. This has become obvious by
Eq.(\ref{2+E_correction_to_R2}): as $\xi$ goes to infinity the least action
principle singles out the manifolds which obey (\ref{Bianchi_identities}),
and segregates the fluctuations in the orthogonal $\,\epsilon\,$ dimensions.
This situation is typical for theories which cannot be continued
self-consistently beyond the number of dimensions they are defined in.
Basically, the effect is not without physical significance: it is a clear
witness of anomaly (see, e.g., an analogous discussion in the two-dimensional
Wess-Zumino-Witten model, \cite{Bos}).

\section{Quantum Reparametrizations and  \protect\\
Asymptotic Freedom in Dilaton Gravity}

The same construction as discussed in the previous section, may be employed
in a conventional (second-order in derivatives) dilaton gravity, which is a
straightforward generalization of the Jackiw-Teitelboim model. With the
help of conformal rescalings of the metric and general transformations of
the scalar field, \cite{Banks_OLoughlin,Russo_Tseytlin},
the action of the most general model may be always written in the form:
\begin{equation}
S=-\int\!d^2x\,\sqrt{g}\left[{1\over2}g^{\mu\nu}\partial_\mu\Phi\partial_\nu
                              \Phi+\Phi R+V(\Phi)\right] \ ,
\label{canonical_form_of_dilaton_gravity}
\end{equation}
so that all the arbitrariness resides in the form of the potential function
$V(\Phi)\,$.

The field transformations cannot remove a unique feature of the dilaton
gravity: the direct dilaton-curvature coupling. However, the coefficients
in (\ref{canonical_form_of_dilaton_gravity}) are subject to change. In
particular, an apparent kinetic term, $(\partial\Phi)^2$, may be set to zero
by an appropriate conformal rescaling of the metric. This should not disturb
us because it is the $\Phi R$-term that carries the genuine (mixed) kinetic
matrix for both the conformal mode and the dilaton. Diagonalizing this
kinetic matrix one finds that the signs of the eigenmodes are opposite so
that the model has zero dynamical degrees of freedom on shell,
\cite{Russo_Tseytlin}.

Within the background field treatment, the action
(\ref{canonical_form_of_dilaton_gravity}) may be supplemented by,
\cite{Kantowski_Marzban},
\begin{equation}
\delta S=\xi\int\!d^2x\,\sqrt{g}\Phi\bar{h}^{\mu\nu}\left[
          \widetilde{R}_{\mu\nu}-{1\over2}\widetilde{R}\widetilde{g}_{\mu\nu}
          \right] \ .
\label{KM-term}
\end{equation}
At one loop, the expression in square brackets should be expanded to the
first order in quantum fluctuations $\{\varphi;\,h;\,\bar{h}_{\mu\nu}\}\,$.
In the simplest covariant gauge,
\begin{equation}
\chi^\mu={1\over(1+\xi)\Phi}\nabla^\mu\varphi-\nabla_\nu\bar{h}^{\mu\nu} \ ,
\end{equation}
\begin{equation}
\widehat{C}_{\mu\nu}=-(1+\xi)\Phi g_{\mu\nu} \ ,
\end{equation}
the correspondent determinant is easily done with the Schwinger-DeWitt
technique yielding
\begin{equation}
{\Gamma_{div}}=-{1\over4\pi\epsilon}\int\!d^2x\,\sqrt{g}\,\Biggl[ 2R+V'+
                                                                  {1\over
                      (1+\xi)\Phi}V-{1\over(1+\xi)\Phi}\Delta\Phi\Biggr] \ ,
\end{equation}
modulo total derivatives of the dilaton. By using the equations of motion for
the background fields,
\begin{equation}
{1\over\sqrt{g}}{\delta S\over\delta\Phi}\equiv\Delta\Phi-R-V'=0 \ , \qquad
{1\over\sqrt{g}}g_{\mu\nu}{\delta S\over\delta g_{\mu\nu}}\equiv\Delta\Phi-V
                                                                      =0 \ ,
\end{equation}
$\Gamma_{div}$ can be put into the form,
\cite{Odintsov_Shapiro,Russo_Tseytlin,Kantowski_Marzban}:
\begin{equation}
{\Gamma_{on\ shell}}={1\over4\pi\epsilon}\int\!d^2x\,\sqrt{g}\,V' \ .
\label{Gamma_div_on-shell}
\end{equation}
The standard renormalization routine gives the following class of
ultraviolet-finite potentials:
\begin{equation}
V(\Phi)=b\exp\left(\alpha\Phi\right) \ ,
\end{equation}
where $b$ and $\alpha$ are arbitrary constants. To understand the
arbitrariness of the exponent $\alpha$ one notes that the second equation of
motion (sometimes termed the ``classical conformal anomaly'') amounts to a
statement that the operator $V(\Phi)$ is a total derivative on shell,
\cite{Kantowski_Marzban},
thus one can add an arbitrary amount of $\smallint\!d^2x\sqrt{g}V$ to
Eq.(\ref{Gamma_div_on-shell}).

The reference to $\xi$ has disappeared from the final on-shell expression, as
it should be for a parametrization-dependent quantity. Moreover, an explicit
evaluation of the determinant reveals that the intermediate expressions in
the heat kernel expansion are regular functions of $\xi$ as long as
$\xi\ne-1\,$. Upon properly accounting for the contributions to the
path-integral measure, all the configuration space matrices which enter the
relevant Seeley-Gilkey coefficient, \cite{Barvinsky-Vilkovisky:review},
remain bounded when $\xi$ grows indefinitely. The limit $\xi\to\infty\,$
returns us to the conformal gauge, up to an on-shell surface term \
$(8\pi\epsilon)^{-1}\!\smallint\!d^2x\sqrt{g}V\,$. We use this circumstance
below to fix the overall scale of the counterterms.

Again, we can re-formulate the source of the $\xi$-dependence introducing the
following piece
\begin{equation}
{\cal A}=-2\xi\mu^{2\epsilon}\int\!d^dx\,\sqrt{g}\,\Phi R \ , \qquad
d=2(1+\epsilon) \ ,
\label{2+E_correction_to_dilaton_gravity}
\end{equation}
and studying $2+\epsilon$-dimensional disturbances of the two-dimensional
background. However, such an approach has an important setback: $\,{\cal A}\,$
is no more zero at $\,\epsilon=0\,$. This crucial difference with the case of
the $R^2$-gravity has to do with the conformal structures of the pertinent
terms in the both models. In fact, there can be no new geometrical term with
the background dimensionality equal to $\,d\,$. If one insists on having the
term $\,{\cal A}\,$ as an addition to
(\ref{canonical_form_of_dilaton_gravity}), then the first equation of motion
changes to:
\begin{equation}
{1\over\sqrt{g}}{\delta(S+{\cal A})\over\delta\Phi}\equiv\Delta\Phi-(1+2\xi)R
                                                    -V'=0 \ ,
\label{modified_eq_of_motion}
\end{equation}
while the equation for the classical conformal anomaly persists. The
reduction of $\Gamma_{div}\,$ with the help of
Eq.(\ref{modified_eq_of_motion}) leads to penetration of the $\xi$-dependence
into the topological divergence, which reminds us of the presence of the
conformal anomaly:
\begin{equation}
\Gamma_{modified}={1-2\xi\over4\pi\epsilon}\int\!d^2x\,\sqrt{g}\,R \ ,
\label{Gamma_div_modified}
\end{equation}
where we have flipped the sign of $\epsilon$ for consistency. Now, there
appears a point in the parameter space, $\xi=1/2\,$, where the on-shell
divergences vanish. This fact may lead us to the following speculation.

As the inclusion of $\delta S^{(2)}$ only affects the
$\bar{h}_{\mu\nu}\bar{h}_{\alpha\beta}$-sector, the two $\Phi R$ terms, in
(\ref{canonical_form_of_dilaton_gravity}) and in
(\ref{2+E_correction_to_dilaton_gravity}), may be thought of as describing
in-the-surface and out-of-the-surface quantum fluctuations, respectively.
When the effect of the latter becomes significant (i.e., well beyond two
dimensions) one can expect that a non-trivial fixed point of the
gravitational beta-function exists. This is already seen in the leading order
of the $\epsilon$-expansion. Replacing the quantity $25-c$ with $24-c$,
\cite{Chamseddine,Russo_Tseytlin},
where $\,c\,$ is the Virasoro central charge for the conformal
matter fields, in the pertinent expressions of
Ref.\cite{Einstein_gravity_in_d=2+E},
one finds the following gravitational beta-function \,($c=0$):
\begin{equation}
\beta(\kappa)\equiv\mu{\partial\kappa\over\partial\mu}=\epsilon\kappa-
                                                             \kappa^3/\pi \ .
\label{beta-function}
\end{equation}
In addition to the trivial fixed point $\kappa=0\,$, there exists
another one,
\begin{equation}
\kappa_c^2=\pi\epsilon \ , \qquad \beta(\kappa_c)=0 \ , \quad
\beta'(\kappa_c)<0 \ .
\end{equation}

The simplest way to see how this obtains in our setting is to take
$\xi\to\infty\,$, which specifies the rigid scale%
\footnote{This procedure is mandatory as emphasized in the first paper of
Ref.\cite{Einstein_gravity_in_d=2+E} because there is no default scale in the
pure Einstein action.}
in the underlying $d=2$ theory, and to freeze the dilaton field at its
natural value $\,\langle\Phi\rangle=1/(2\kappa^2)\,$. Then we have
\begin{equation}
{1\over2\kappa_0^2}=\mu^{2\epsilon}\left({1\over2\kappa^2}-{1\over2\pi
                                                      \epsilon}\right) \ ,
\end{equation}
where $\kappa_0$ is the bare Einstein constant, which leads directly to
Eq.(\ref{beta-function}).

Dropped from Eq.(\ref{Gamma_div_modified}) are the convergent terms: those
contribute ${\cal O}(\kappa^3\epsilon)$ to the beta-function
(\ref{beta-function}). As the loop expansion is justified for small values
of $\,\kappa\,$, the omitted contributions are indeed negligible whilst the
two remaining terms may be on the same order of magnitude,
$\,\kappa^2\simeq\epsilon\,$. The latter is manifest for the ultraviolet
stable point $\,\kappa_c\,$.

\section{Conclusions}

In this paper, we have studied quantum reparametrizations in
two-dimen{\-}sional models of gravity. We have pursued the point of view
that a dimensional extension $d=2n\to2(n+\epsilon)$ of a geometrical theory
is encoded in the parametrization structure of the model formulated in the
basic number $2n$ of dimensions. This has been demonstrated on a
comparatively simple example of the two-dimensional $R^2$-gravity.
The related discussion for the (conventional) dilaton gravity has also been
presented.

We have provided some support for a popular speculation that the
strong-coupling gravity belongs to a different universality class from its
weak coupling version. In the dilaton gravity, we have found an ultraviolet
fixed point for the Einstein constant flow to the leading order in the
$\epsilon$-expansion around two dimensions. It would be interesting to
couple matter fields to the action (\ref{canonical_form_of_dilaton_gravity})
and to find the relevant operators at the fixed point $\kappa_c\,$. We must
admit, however, that although the dilaton gravity is as legitimate as the
Einstein theory at $d=2$ they definitely have different behaviors in higher
dimensions.

Technically, we have shown how to proceed with the Schwinger-DeWitt
technique in the two-dimensional $R^2$-gravity where the highest-derivative
term in the one-loop determinant is degenerate within the linear metric
parametrization and no operator squaring \cite{Barvinsky-Vilkovisky:review}
can help. The construction is straightforwardly generalized to a much more
complicated case of a scalar-tensor $R^2$-gravity
\cite{our_R2_dilaton_gravity}.
We believe that our approach may be useful in other models as well.

Our final remark concerns another possible interpretation of
Eq.(\ref{KM-term}). Note that the background field $\Phi$ is an artifact of
a specific realization: it simply sets the scale. Contrary to that, the
``graviton'' field $\bar{h}_{\mu\nu}\,$ may be viewed as a Lagrange
multiplier that enforces the two-dimensional Bianchi identities. When
Eq.(\ref{Bianchi_identities}) holds  the Lagrange multiplier is unimportant;
conversely, when the auxiliary field is introduced the constraint
(\ref{Bianchi_identities}) may be relaxed. This observation might be the
first step towards the dual description of low-dimensional gravity.

\bigskip
\bigskip

{\bf Acknowledgements}\medskip\par\noindent
The author is grateful to S.D. Odintsov, D.P. Sorokin, D.V. Volkov, and
A.A. Zheltukhin for fruitful discussions at different stages of the work,
and to V.A. Miransky for a careful reading of the manuscript. The research
was supported by the International Science Foundation.


\end{document}